# An Analog Neural Network Computing Engine using CMOS-Compatible Charge-Trap-Transistor (CTT)

Yuan Du*, *Member, IEEE,* Li Du*, *Member, IEEE,* Xuefeng Gu, Jieqiong Du, X. Shawn Wang, *Student Member, IEEE,* Boyu Hu, Mingzhe Jiang, Xiaoliang Chen, Junjie Su, Subramanian S. Iyer, *Fellow, IEEE,* and Mau-Chung Frank Chang, *Life Fellow, IEEE*

***Abstract*— An analog neural network computing engine based on CMOS-compatible charge-trap transistor (CTT) is proposed in this paper. CTT devices are used as analog multipliers. Compared to digital multipliers, CTT-based analog multiplier shows significant area and power reduction. The proposed computing engine is composed of a scalable CTT multiplier array and energy efficient analog-digital interfaces. By implementing the sequential analog fabric (SAF), the engine's mixed-signal interfaces are simplified and hardware overhead remains constant regardless of the size of the array. A proof-of-concept 784 by 784 CTT computing engine is implemented using TSMC 28nm CMOS technology and occupies 0.68mm$^2$. The simulated performance achieves 76.8 TOPS (8-bit) with 500 MHz clock frequency and consumes 14.8 mW. As an example, we utilize this computing engine to address a classic pattern recognition problem -- classifying handwritten digits on MNIST database and obtained a performance comparable to state-of-the-art fully connected neural networks using 8-bit fixed-point resolution.**

## I. Introduction

DEEP learning using convolutional and fully connected neural networks has achieved unprecedented accuracy on many modern artificial intelligence (AI) applications, such as image, voice, and DNA pattern detection and recognition [1-10]. However, one of the major problems that hinder its commercial feasibility is that neural networks require great computational resources and memory bandwidth even for very simple tasks. The insufficient ability of modern computing platforms to deliver simultaneously energy-efficient and high performance leads to a gap between supply and demand. Unfortunately, with current Boolean logic and Complementary Metal Oxide Semiconductor (CMOS)-based computing platforms, the gap keeps growing mainly due to limitations in both devices and architectures. State-of-the-art digital computation processors such as CPUs, GPUs or DSPs in the system-on-chip (SoC) systems [11-13] cannot meet the required computational throughput within the power and cost constraints in many practical applications. Most of the modern computational processors are all implemented on Von-Neumann architecture where the processor and main memory are separated. However, as transistor technology scaling reaches its physical limits, the computational throughput using current architectures will inevitably saturate [14]. Furthermore, the extensive processor-memory data movements have become the dominated energy consumption [39].

To improve the performance and energy efficiency of the neuromorphic system, several analog computing architectures have been reported to significantly reduce the power-hungry data transfer and avoid the memory wall [15-16, 34-35, 37-38]. By integrating computation into memory, these architectures show great potential to achieve high performance and efficiency. For example, PRIME [34] employing ReRAM-based memory array for neuromorphic computation shows a potential of 2~3 orders improvement in terms of system performance and energy efficiency. ISAAC [35] proposes a crossbar-based pipeline architecture and data encoding techniques to reduce data converters' complexity. Eight memristors each storing 2-bit is used to support 16-bits resolution and multi-bit data are calculated in a bit-to-bit sequential fashion to reduce DAC/ADC resolution. However, ISAAC does not provide circuit level implementation and simulations. Even though the simulation results from PRIME and ISAAC show great potential, the task of fabricating large memristor arrays is still very challenging. In addition, various analog devices have also been demonstrated for these applications [17-18]. However, these devices require the introduction of new materials or extra manufacture processes, which are not currently supported by major CMOS foundries. Thus, they cannot be embedded into the large volume commercial CMOS chips.

Recently, charge-trap transistors (CTTs) were demonstrated to be used as digital memory devices in [19-20] with error-proof trapping and de-trapping algorithm. Compared to other analog devices based on charge-trapping phenomena such as floating-gate transistors [21], transistors with an organic gate dielectric [22], and carbon nanotube transistors [23], CTTs are fully CMOS-compatible in terms of process, operating voltage and manufacturing maturity. Even though the charge-trapping phenomenon in a transistor with high-k-metal gate has reliability concerns, causing bias temperature instability, it was recently discovered that with a drain bias during the charge-trapping process, many more carriers can be trapped in the gate dielectric stably, and more than 90% of the trapped charge can be retained after 10 years even when the device is baked at 85 °C [24]. In addition to being reliable, CTT arrays are also demonstrated to be high-density and scale well with current commercial CMOS/FinFET technology [19]. More interestingly, an analog synapse vector was demonstrated to execute unsupervised learning computation in [25]. However, the demonstrated analog synapse vector only includes nine neurons, which do not suffice to perform any practical neural network computation; it did not consider the analog and digital interfaces, which will be the main energy consumption in memristor-based neuromorphic computing applications.

*These authors contributed equally to this work.
Y. Du, L. Du, X. Gu, J. Du, X.S. Wang, B. Hu, M.C. F. Chang are with the High Speed Electronics Lab (HSEL), University of California, Los Angeles, CA, 420 Westwood Plaza, Los Angeles, CA 90095
X. Gu, S. S. Iyer are with Center for Heterogeneous Integration and Performance Scaling (CHIPS), Electrical Engineering Department, University of California, Los Angeles , CA 90095
X. Chen is with University of California, Irvine, CA 92697
M.-C. F. Chang is also with National Chiao Tung University, No. 1001, Daxue Rd, Hsinchu, Taiwan
Y. Du, L. Du, M. Jiang, J. Su are also with Kneron, Inc, San Diego, CA 92121

In this paper, we propose an analog computing engine based on charge-trap transistors (CTTs). The proposed computing engine is composed of a 784 by 784 CTT analog multipliers and achieves 100x power and area reduction compared with regular Boolean CMOS-based digital computation platforms. By implementing a sequential analog fabric (SAF), the mixed-signal analog-to-digital interfaces are simplified and it only requires an 8-bit analog-to-digital converter (ADC) in the system. The top-level system architecture is shown in Fig. 1, which will be discussed in detail in Section IV. The main contribution of this paper includes:

(1) An 8-bit 784x784 parallel fully connected neural network (FCNN) analog computing engine, using CTT-based analog multipliers, is designed and achieves dramatic area and power reduction compared to the conventional Boolean CMOS-based computing engine.

(2) A system-level architecture with area and energy efficient analog-digital interfaces is developed to flexibly store, calibrate or re-process inter-layer partial calculation results to guarantee analog computation accuracy.

(3) A detailed circuit implementation of sequential analog fabric (SAF) that is adopted from the bit-to-bit computation idea from ISAAC and dedicated for our CTT-based computing engine. It simplifies the interfaces between analog and digital domain, by removing the required digital-to-analog converter (DAC) and enabling the parallel computation of multiple neurons.

(4) A practical application, handwritten digit recognition using different configurations of multilayer neural network structure, is well simulated and analyzed based on single device experimental data over MNIST dataset.

(5) A number-of-bit resolution requirement study is performed, showing that the resolution of the 8-bit fixed-point data format is good enough to achieve similar performance, compared with that of the 32-bit floating-point data format (difference less than 2%).

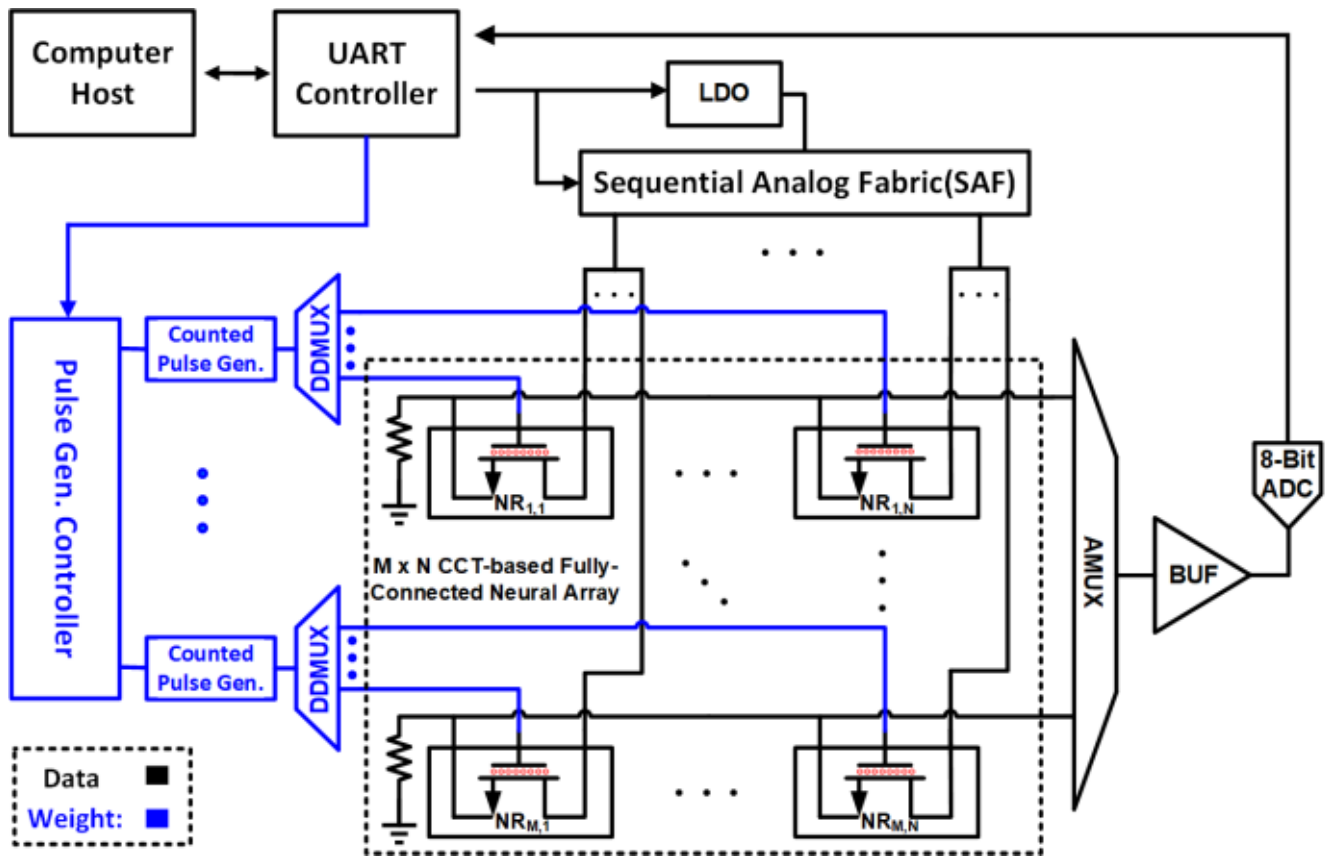

Fig. 1. Top-level system architecture of the proposed analog computing engine, including a CTT array, mixed-signal interfaces including a tunable low-dropout regulator (LDO), an analog-to-digital converter (ADC), and a sequential analog fabrics (SAF)

This paper is organized as follows: we first introduce the basics of CTT device physical principles and discuss how to use the CTT device to perform analog multiplication in Section II. In Section III, system-level challenges and considerations are described. The detailed building block designs, operations, and sample experiment results are reported in Section IV, V and VI, respectively. Finally, the conclusion is drawn in Section VII.

## II. CHARGE-TRAP-TRANSISTOR DEVICE INTRODUCTION

### A. *CTT Basics*

Charge-trapping phenomenon is well-known in devices using flash memory [27]. However, it is not favored for high-performance logic or low-cost foundry technologies because it may require additional masks, higher process complexity, or its operating voltage is incompatible. On the other hand, a fully logic-compatible CTT is reported in 22 nm planar and 14 nm FinFET technology that do not add process complexity or masks in [28]. With enhanced and stabilized charge-trapping behavior, the CTTs show promise to be exploited as basic analog computing elements for neural networks.

N-type CTTs with an interfacial layer (IFL) $SiO_2$ followed by an $HfSiO_x$ layer as the gate dielectric is used in [28], which is a common recipe in state-of-the-art CMOS technologies. It should be noted that, although it is demonstrated only on planar SOI devices, the mechanisms apply to bulk substrates of FinFETs as well.

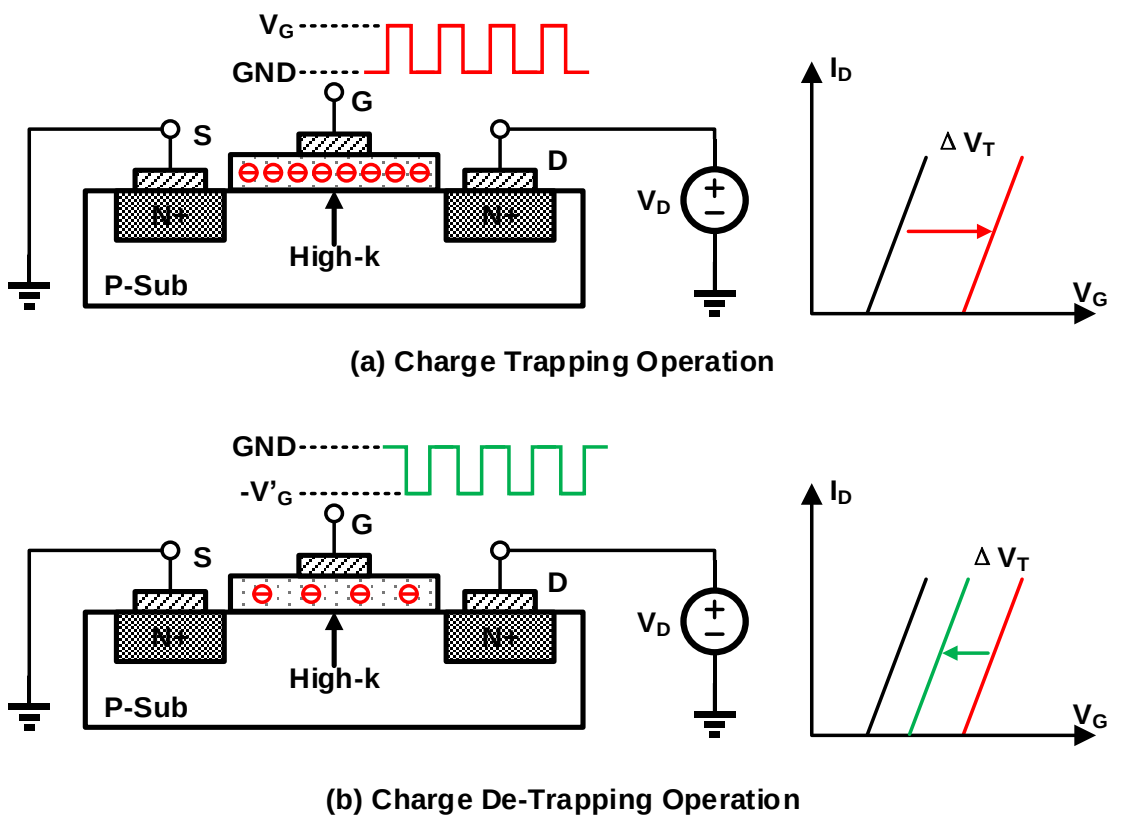

Fig. 2. A schematic showing the basic operation of CTT device (equally applicable to FinFET based CTTs): (a) Charge Trapping Operation; (b) Charge De-Trapping Operation

A schematic of the basic operation of a CTT device is depicted in Fig. 2. The device threshold voltage $V_T$ is modulated by the amount of charge trapped in the gate dielectric of the transistor. To understand the dynamic behavior of charge trapping, we first applied a pulse train at the gate terminal of a CTT device, and then measured the $V_T$ change as a function of the applied pulse number, as shown in Fig. 2(a). CTT devices can be programmed by applying µs-long trapping (positive) and de-trapping (negative) pulses on the gate to modify the threshold voltage of the transistor. For the examples in [28], 2V pulses were used during charge trapping operation with 1.3V drain voltage; during charge de-trapping operation, -1.3 V pulses are used with 0V drain voltage. Programming efficiency is highest at the beginning of the program operation and reduces with increasing programming time as more and more of the available electron traps are filled. A drain bias enhances and stabilizes the charge-trapping process. The trapped charge dissipates very slowly (> 8 years at 85 °C), allowing CTTs to be used for embedded nonvolatile memory [19]. Furthermore, because CTTs are built based-on commercial standard NMOS transistors, the process variation is well controlled with high yield rate. It is more mature to use a large amount of CTTs to build large-scale analog computing engines, compared with other emerging analog computing technologies. More attractively, very low energy consumption per synaptic operation is reported at pico-joule level [25].

## *B. CTT-based Multiplication*

For most of the neuromorphic networks, the training and inference operations rely heavily on vector or matrix multiplication in both feedforward and error back-propagation computation. Fig. 3 (a) shows an M-by-N fully connected neural network, also known as a fully connected layer. The input data $X_i$ and the output results $Y_j$ are connected by a weighted M by N matrix. Each output $Y_j$ is determined by:

$$Y_j = \sigma\left(\sum_{i=1}^{M} X_i W_{i,j}\right), \qquad (1)$$

where $W_{i,j}$ is the synaptic weight coefficient between the input neuron i (i=1…M) and the output neuron j (j=1…N) and σ is a non-linear activation function like sigmoid or ReLU.

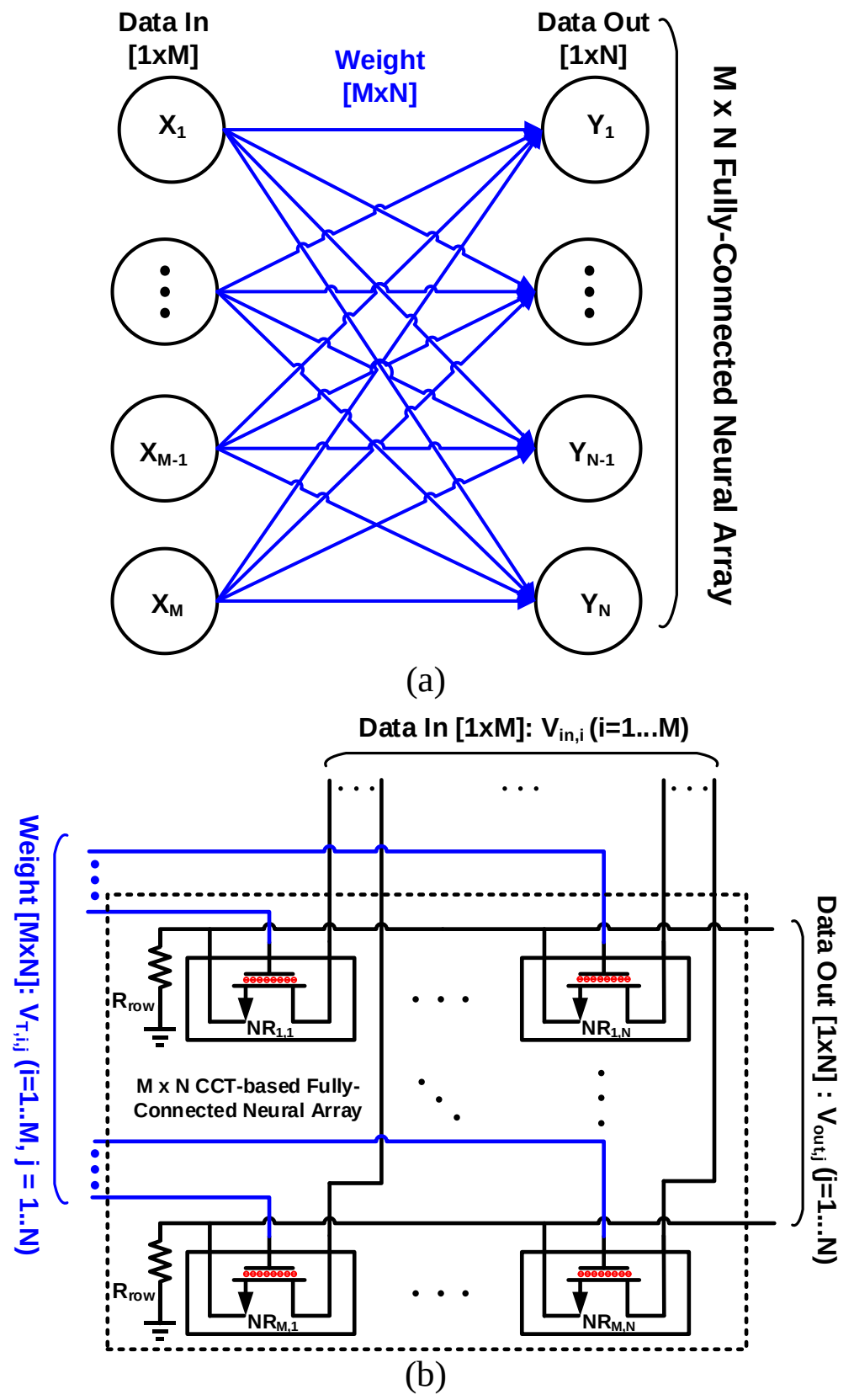

Fig. 3. (a) Fully-connected neural array; (b) CTT multiplication array (Op-amp buffers are not shown)

In a conventional all-digital implementation, it requires a lot of data movement for synaptic weights reading for the dot product computation, which is power-hungry. However, a CTT array can easily fulfill this task: 1) the weight coefficients are stored locally through programming the CTT threshold voltages, and 2) multiplication and accumulation operations are performed through an M-by-N crossbar structure shown in Fig. 3 (b). In the real implementation, there is an operational amplifier based buffer in each row to maintain a relative small change due to different current values flowing through source resistors.

In the CTT array, a device cell ($NR_{i,j}$) is a synapse, connecting one input data $X_i$ to one output result $Y_j$. The device $NR_{i,j}$ stores the synaptic weight ($W_{i,j}$) through threshold voltages $V_T$ of the device $NR_{i,j}$. The device also accomplishes multiplication operation $W_{i,j} \cdot X_i$ when operating in triode region and taking input data through the drain-source voltage $V_{DS}$. When biased in triode region where $V_{DS}$ is smaller than the gate-source voltage $V_{GS}$ by one $V_T$, the drain-source current $I_{DS}$ of the device approximates a linear function of the product of

$V_{DS}$ and $V_T$ as equation (2) shows [36]:

$$I_{DS}(V_{DS}, V_T) = \frac{1}{2} k_n \frac{W}{L} \left[2V_{DS}(V_{GS} - V_T) - V_{DS}{}^2\right] \approx k_n \frac{W}{L} V_{DS}(V_{GS} - V_T),$$
$$\text{when } V_{DS} < V_{GS} - V_T, \quad (2)$$

where $K_n$ is the process transconductance parameter and W and L represent the width and length of the transistor, respectively.

The $V_T$ of each transistor in CTT array can be programmed by varying the duration of the positive/trapping or negative/de-trapping pulses trains. Due to the fast-reading and slow-writing nature of CTT $V_T$ [19], it is unique in storing weights in the CTT threshold voltages and providing weights during the neural network inference mode when the weights are fixed after being programmed from the pre-trained model. While $V_T$ stores the weight value, the input data value could be converted to $V_{DS}$ through a voltage reference source. The gate-source voltage $V_{GS}$, on the other hand, is constant during multiplication operation and will be calibrated out eventually.

A resistor $R_{row}$ collects all the drain-source currents from the synapses in the row and then performs summation. If the input data values were available at the same time, all the output data would be ready within one clock cycle. Through the resistor, current signals are converted to voltage signals. The voltages across the row resistors can be calculated by the following equations (3), (4), and (5).

$$V_{out,j\,(j=1\ldots N)} = R_{row} \cdot \sum_{i=1}^{M} I_{D,i,j} \quad (3)$$

$$V_{out,j(j=1\ldots N)} = R_{row} \cdot k_n \cdot \frac{W}{L} \sum_{i=1}^{M} V_{DS,i,j} \cdot (V_{GS} - V_{T,i,j}) \quad (4)$$

$$V_{out,j(j=1\ldots N)} = R_{row} \cdot k_n \cdot \frac{W}{L} \sum_{i=1}^{M} V_{DS,i,j} \cdot V_{T,i,j} - f(V_{DS,i,j}) \quad (5)$$

Here, $V_{out,j}$ represents the output of $Y_j$ neural cell at Row j, $V_{DS,i,j}$ is transferred from input image pixel value, $V_{T,i,j}$ is programmed by the pulse number based on pre-trained model $W_{i,j}$ value, and $R_{row}$ is the resistance of the summing resistor. As is shown in (5), the right side of the equation is separated into two terms. The first term is the wanted multiplication of $V_{DS}$ and $V_T$ while the second term is an unwanted input-data-dependent offset. Fortunately, the input data is known in the system and the offset could be easily calibrated out after the analog-to-digital converter in the digital domain.

## III. System-level Architecture

### A. System-level Considerations

Section II introduces the fundamentals of CTT devices and shows how it is possible to use CTT devices to compose an array to compute vector or matrix multiplication effectively in parallel. In this section, system-level considerations are discussed.

To compare the energy and area efficiency of CTT-based computations with conventional digital domain computations [26], the energy consumption per multiplication operation for a CTT device is one order lower than its 32-bit floating-point digital counterpart. One 8-bit MAC requires approximately 100~400 transistors while one CTT-based MAC only needs a single CTT transistor. However, an analog-digital interface conversion circuitry is required for analog computing which adds to the system's overall power consumption.

Although CTT technique show promise to achieve low-power high-performance matrix computation in parallel, there are three important problems needed to be solved to utilize the CTT-based computation technique into practice; we will focus on solving (1) and (2) in this paper:

(1) An efficient interface between analog and digital domain that enables fast and easy data format transfer between the analog and digital domains.
(2) A scalable and reconfigurable array that computes parallel multiple neuron values simultaneously.
(3) A robust training and inference algorithm to tolerate nonlinearities, process variations, and other computing uncertainties.

### B. Top-level System Architecture

To solve the aforementioned issues, we propose a CTT-based array architecture for an efficient fully-connect layer computation, in Fig. 1, including a 784-by-784 CTT multiplication array, mixed-signal interfaces including a tunable low-dropout regulator (LDO), an analog-to-digital converter (ADC), and a sequential analog fabric (SAF) to assist parallel analog computing.

The size of the multiplication array could be scalable, while the mixed-signal interfaces hardware overhead is almost constant. The intermediate data, which has been converted to the digital domain, could be stored in any type of on-chip/off-chip memory and then applied to another layer to enable multi-layer neural networks. In the proof-of-concept prototype, the inter-layer data will be stored in PC memory through UART interface.

The sequential analog fabric array block is critical to feed multiple drain voltage in parallel using only one voltage reference. A single 8-bit ADC will be used to read out the partial summation results in each row. The designs of key building blocks will be detailed in the next section. A study of the resolution requirement is shown in Section V.

## IV. Building Block Designs and Operations

### A. Design of Key Building Blocks

#### 1) Sequential Analog Fabric

A sequential analog fabric (SAF) shown in Fig. 4 is implemented in the engine to enable parallel analog computations of the multiple neurons. Similar to the bit-to-bit data feeding in [35], when a set of the input neurons are fed into the sequential analog fabric, the fabric first transfers each data's parallel input bits into a sequence. Then each bit of the neurons will control the analog switches in sequence to turn on/off the corresponding CTT multipliers. The computed results of the analog multipliers will be summed at the row resistors and sampled at the ADC input. Each bit computation will take one clock cycle. Different bits' output will be accumulated at the digital domain after the ADC completes the sampling. For our 8-bit data format, it will use eight clock cycles to finish one array's fully connected multiplication. The switch size of the analog fabric is carefully tuned to keep its on-resistance to be less than 20 Ohm, which in this case will not complicate the pre-amplifier design and will not affect the overall computation accuracy.

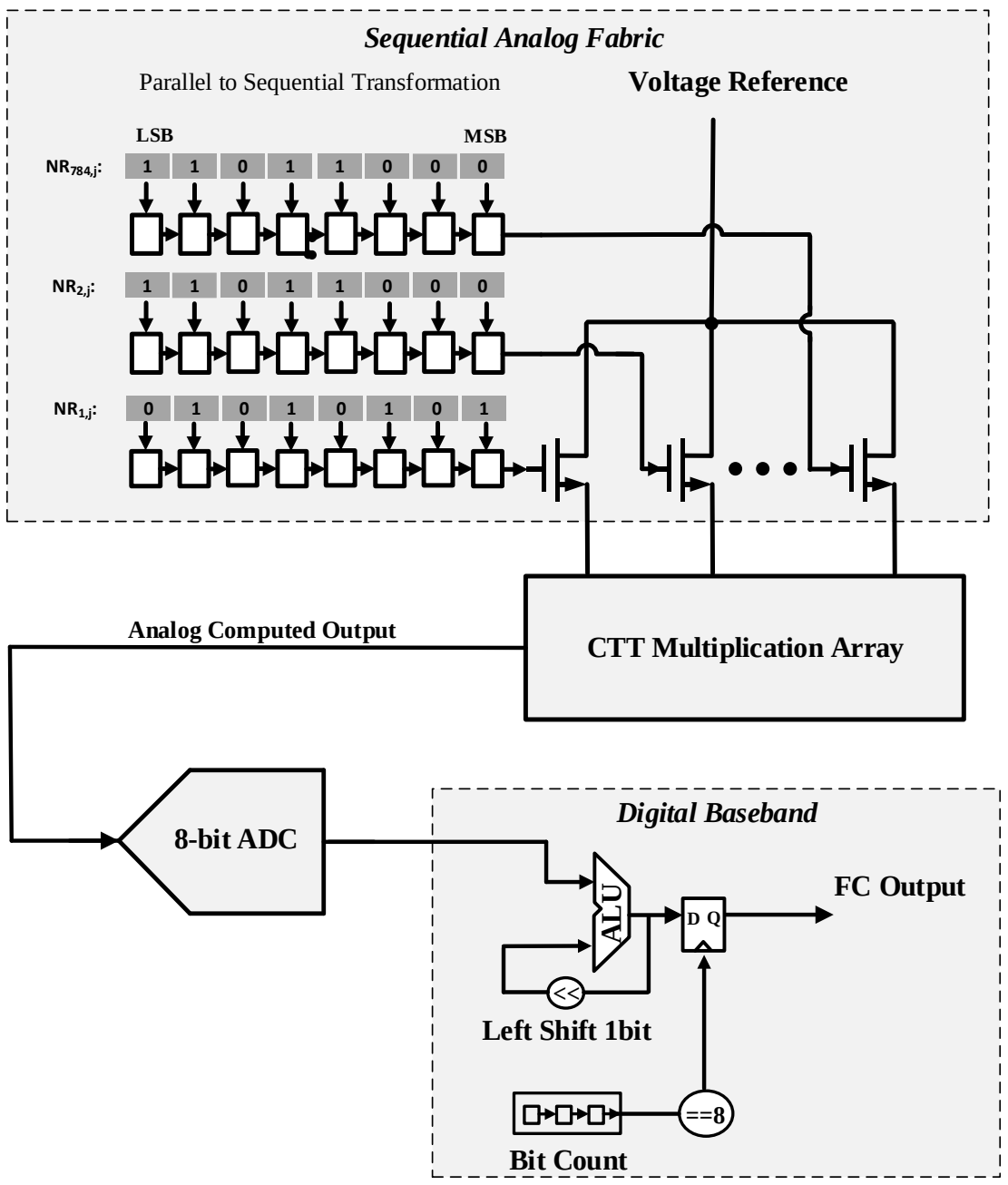

Fig. 4. Schematic design of Sequential Analog Fabric (SAF) block

Since only 1-bit of each neuron will be sent out to the multiplication array, the CTTs' drain node side will be either a fixed voltage or floating. Thus, the nonlinearity introduced by the $V_{DS}$ will become a constant offset in the computation which can be calibrated out. Compared with regular analog computing, no digital-to-analog converter (DAC) is required to generate multi-level input voltage for the CTTs array. In addition, since the applied voltage is constant, the required dynamical range of the sampling ADC is also reduced.

Besides mixed signal interface reduction, the analog fabric also improves the engine performance through enabling parallel neurons data to be fed into the CTT multipliers' array simultaneously. As the input drain voltage to each multiplier is fixed, only a switch is required to turn on/off the multiplier based on the current input bit value.

*2) Analog-to-Digital Converter*

To quantify the computed results of the CTT multiplication array, an 8-bit low power SAR ADC is implemented using an asynchronous architecture in Fig. 5, which achieves better power and speed performance compared with its synchronous structure counterparts and does not require multiple phase-matched ADC clocks to be distributed. The SAR ADC is connected to the amplifier's output to sense the computed analog voltage. To improve the efficiency, the SAR ADC uses a sub-radix [31] and two-capacitor DAC [32] to provide over-range protection to capacitor mismatches and insufficient settling at the expense of one more conversion cycle.

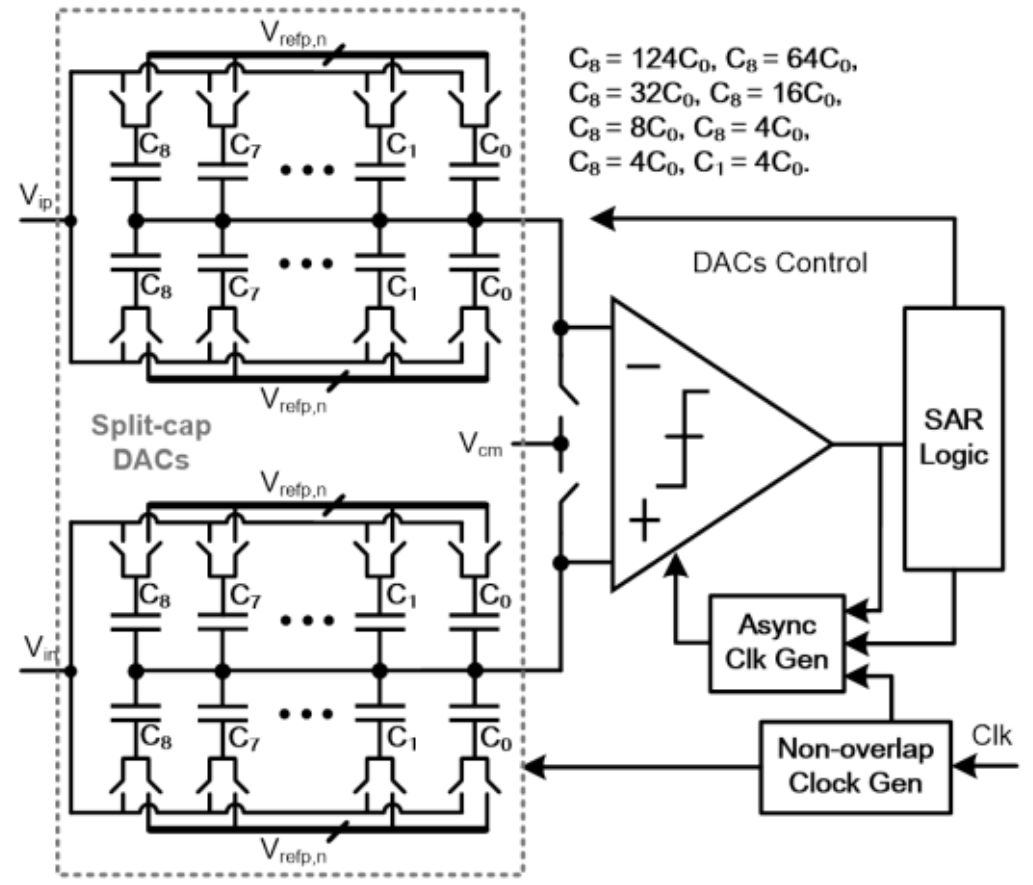

Fig. 5. Schematic design of 8-bit low power SAR ADC block

The comparator in Fig. 6 uses a double-tail latch topology with an integrator ($M_{1P}/M_{1N}$) followed by three parallel differential pairs ($M_{2aP}/M_{2aN}$, $M_{2bP}/M_{2bN}$, $M_{3bP}/M_{3bN}$) and a regenerative latch to accommodate the 1V low supply voltage. The latch-reset differential pairs help to minimize the regeneration time by minimizing the device capacitances. When clk is low, the nodes $d_{ip}$ and $d_{im}$ are reset to supply while the outputs op1 and on1 are discharged to ground. When CLK goes high, $d_{ip}$ and $d_{im}$ begin discharging to the ground while the differential input signal $V_{IP}–V_{IN}$ is being integrated and amplified to $d_{ip}–d_{im}$. When $d_{ip}$ or $d_{im}$ is low enough to turn on $M_{2aN}$ or $M_{2aP}$, regeneration is triggered. A small differential-pair injecting correction current is added at the latch input for offset calibration, instead of a capacitive load of input transistors, because the heavy capacitive load increases the integration time that affects the speed.

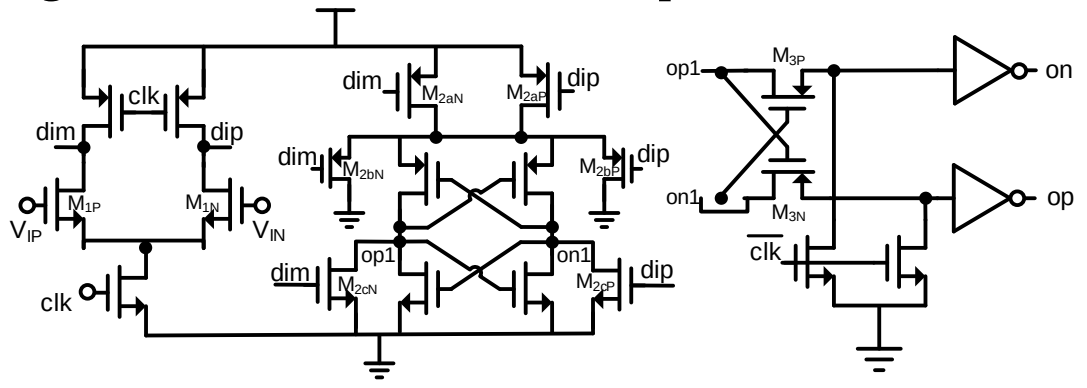

Fig. 6. Schematic design of comparator block

### B. Operation Procedure

The operation of the proposed engine is simple and effective. The pre-trained weight values will be written into the CTT array by counted pulse generators under the control of a pulse generator controller. The weight values are transferred from the digital domain to representing the conductance of CTT devices. For each column, the CTT drains are connected together in order to reduce the number of input port hardware overhead. The drain voltage represents the neuron's value. To enable parallel neurons' computation, each neuron's value is decomposed into several bits and fed into the array in sequence, which will be handled by the SAF block. The first necessary operation in the digital domain after ADC sampling is a sequential accumulation to sum all the decomposed bit components in the SAF and recover complete results including the full resolution. The calculated partial summation of each bit is accumulated together in the digital domain.

Before starting actual computation, a group of calibration data with known input value will be loaded into the CTT array. The correct calculation results have already been stored in the digital domain. The calculation results will be sampled and fed into calibration algorithm. This process is the calibration initialization.

For a 784 x 784 CTT array, the weight writing clock-cycle number of the whole engine is equal to 784 times the longest pulse number. This is because 784-counted pulse generators program the CTT device column by column and the longest pulse number determines how fast to program one column weight. This process might be quite slow and needs an extra error-correction algorithm to maintain weight accuracy. Once programming is done, those values are nonvolatile and the forward propagation or inference speed is fast because of CTT device's slow writing and fast reading features. Consequently, the proposed computing engine is mainly targeting inference computation, rather than training process.

The computation throughput is able to achieve 76832 MACs per clock cycle. Equivalently, it is around 76.8 TOPS per sec with a 500 MHz clock frequency. The detailed procedure flow is shown in Fig. 7.

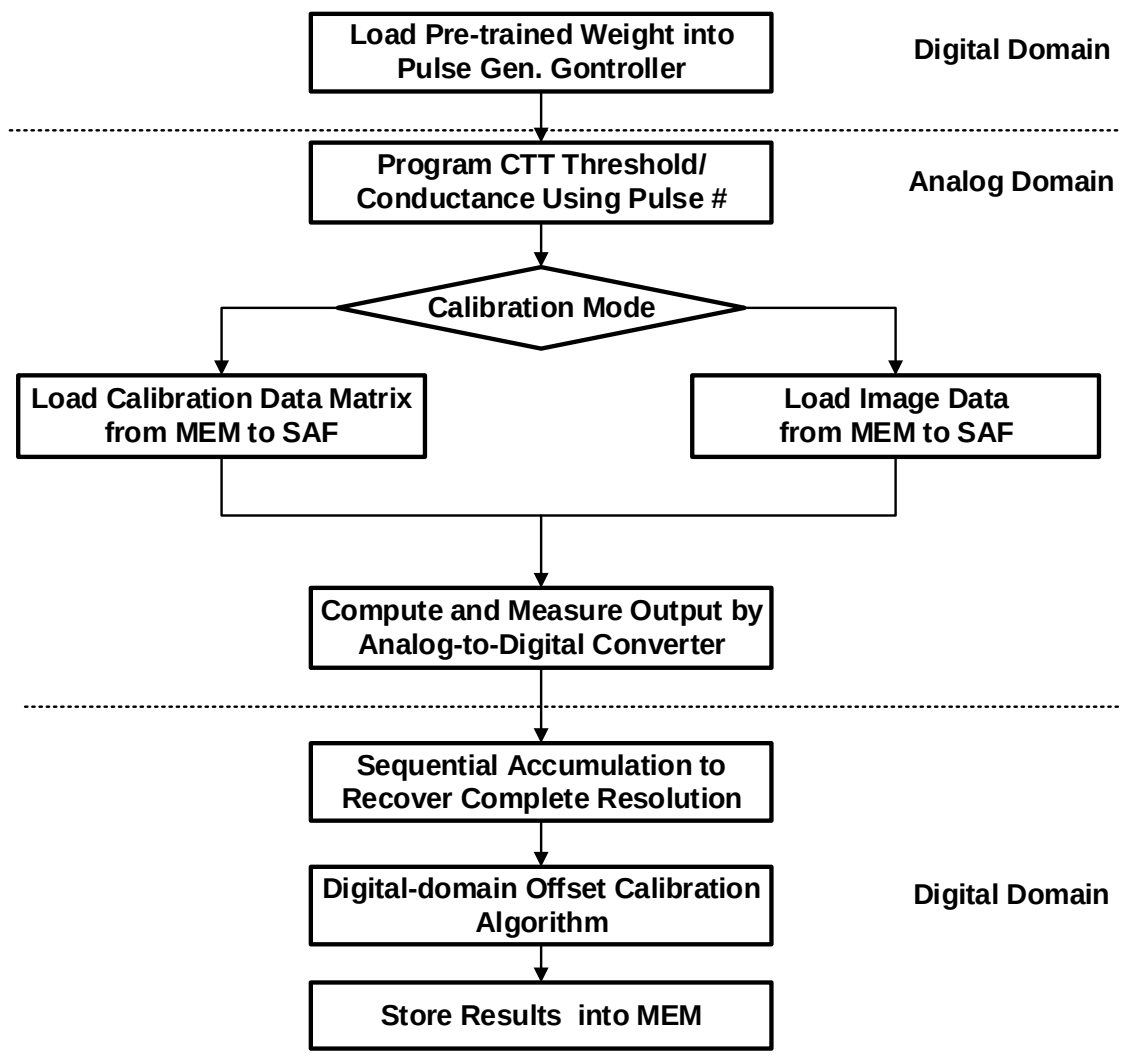

Fig. 7. Commuting engine operation flow chart

In this paper, we designed three different configurations of fully connected neural networks for the handwritten digit recognition problem and use our proposed analog computing engine to compute them. The array element number is chosen based on the 28x28-image size in MNIST database. With 784x784 array size choice, complete fully-connect layer computing could be finished within one operation cycle. The CTT device model comes from the experiment results in [25]. With mixed-signal analog-digital interfaces, the inter-layer partial results could be stored in any type of available memory system. It is necessary because digital-assistant calibration and optimization algorithm could be utilized seamlessly to guarantee analog computing accuracy. In this proof-of-concept prototype, they could be stored in the PC's hard drive conveniently through a low-speed UART interface.

## V. EXAMPLE OF HANDWRITTEN DIGITS CLASSIFICATION

Handwritten digit recognition is an important problem in optical character recognition, and it has been used as a benchmark for theories of pattern recognition and machine learning algorithms for many years. The freely available Modified National Institute of Standards and Technology database (MNIST) of handwritten digits has become the standard for fast-testing machine learning algorithms [29]. Samples of 28x28-pixel images in MNIST are displayed in Fig. 8.

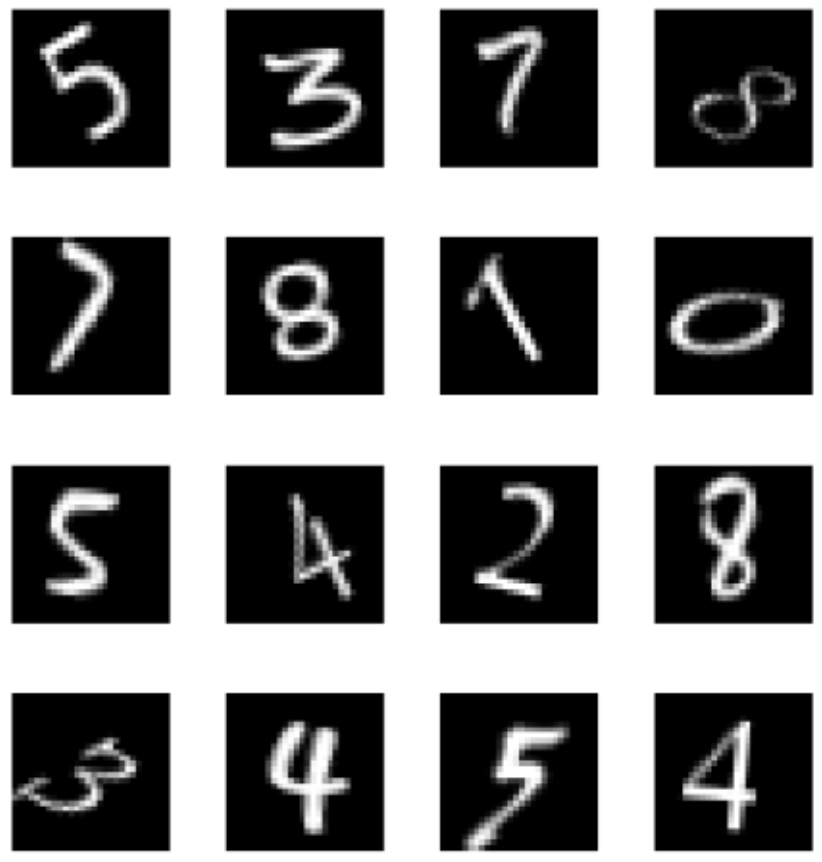

Fig. 8. Samples from MNIST 28 x 28-pixel handwritten digits images

The impact of analog and digital interfaces resolution are studied and the simulation results are shown in Fig. 9. A different number of bits are swept from 1 bit to 16 bits for three different network structures. Case (1): Two layers without hidden layer; Case (2): Three layers with 784 input neural cells, 300 cells in one hidden layer and 10 output cells; Case (3): Four layers with 784 input neural cells, 300 cells in the first hidden layer, 100 cells in the second hidden layer, and 10 output cells. It achieves recognition accuracy 69.8%, 94.2%, and 95.7%, respectively in Case (1), (2), and (3) using 16-bit fixed-point resolution on 10,000 testing images in MNIST database.

In the case of resolution less than 5 bits, too many overflows and underflows exist, which makes accuracy very low for all network configurations. However, in the case of resolution between 6 bits and 16 bits, the recognition accuracies are dramatically improved and comparable to using the 32-bit floating-point data format.

If the 8-bit resolution is chosen as proposed in Fig. 1, more than 94% accuracy can be obtained in Case (2) and (3). In all three cases, the accuracy difference between 32-bit floating-point and 8-bit fixed-point is within 2%. Compared with the 16-bit or 32-bit computation, the 8-bit resolution reduced the hardware overhead significantly at the small cost of accuracy loss.

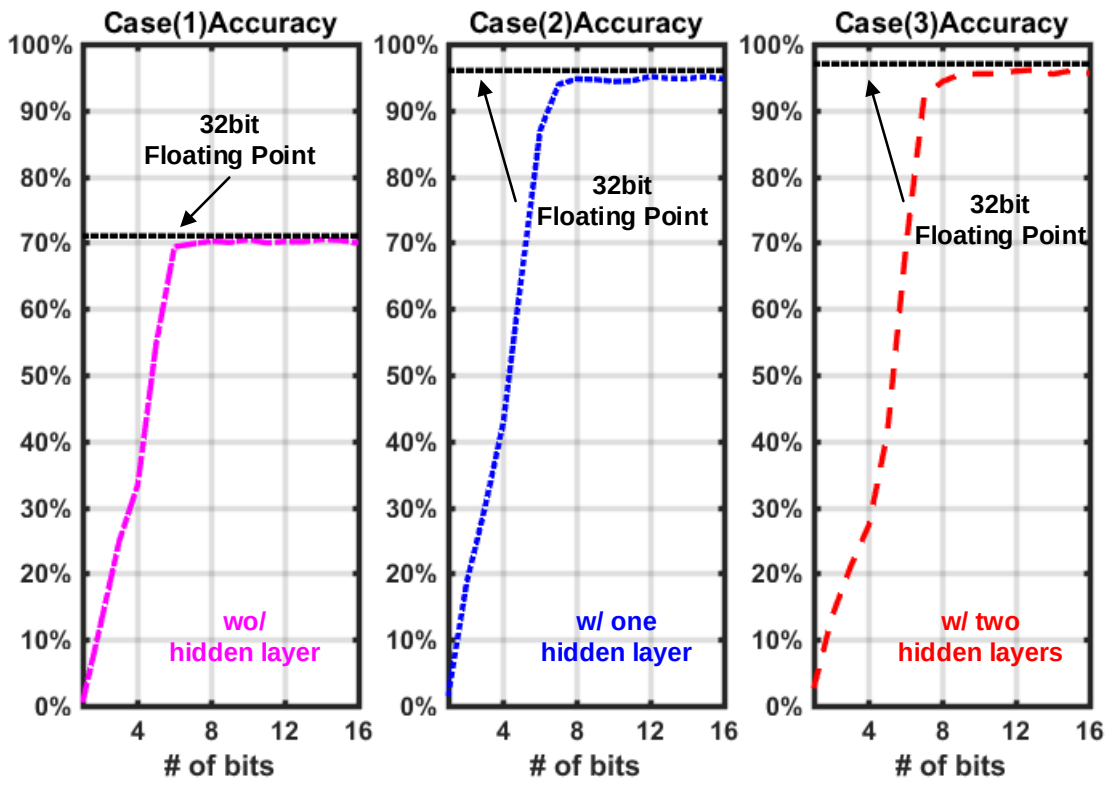

Fig. 9. Accuracy versus resolution of analog-digital interface

## VI. Physical Design and Performance Summary

The analog computing engine is implemented in TSMC 28 nm CMOS HPM standard $V_T$ technology. To evaluate the area, power and critical path of the pulse generator and controller, we developed the register-transfer level (RTL) design in Verilog, and then it is synthesized using Synopsys Design Compiler. We placed and routed the engine using Cadence Innovus. The 8-bit ADC is a silicon-proof IP in the same technology. The dynamic and static power consumption is estimated by Synopsys Prime Time. The other parts are designed and simulated in Cadence Virtuoso. The layout view is shown in Fig. 10. The total core area is 0.68 $mm^2$ and area breakdown is shown in Fig. 11.

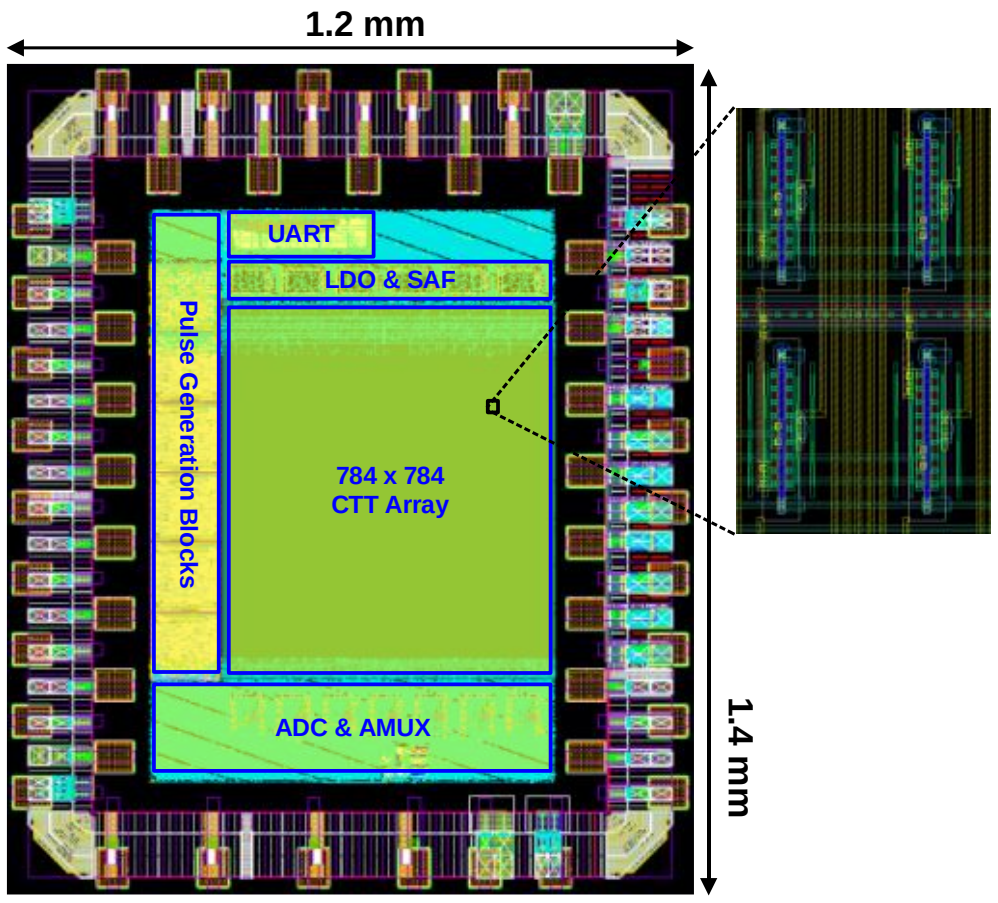

Fig. 10. Layout view in TSMC 28nm CMOS technology

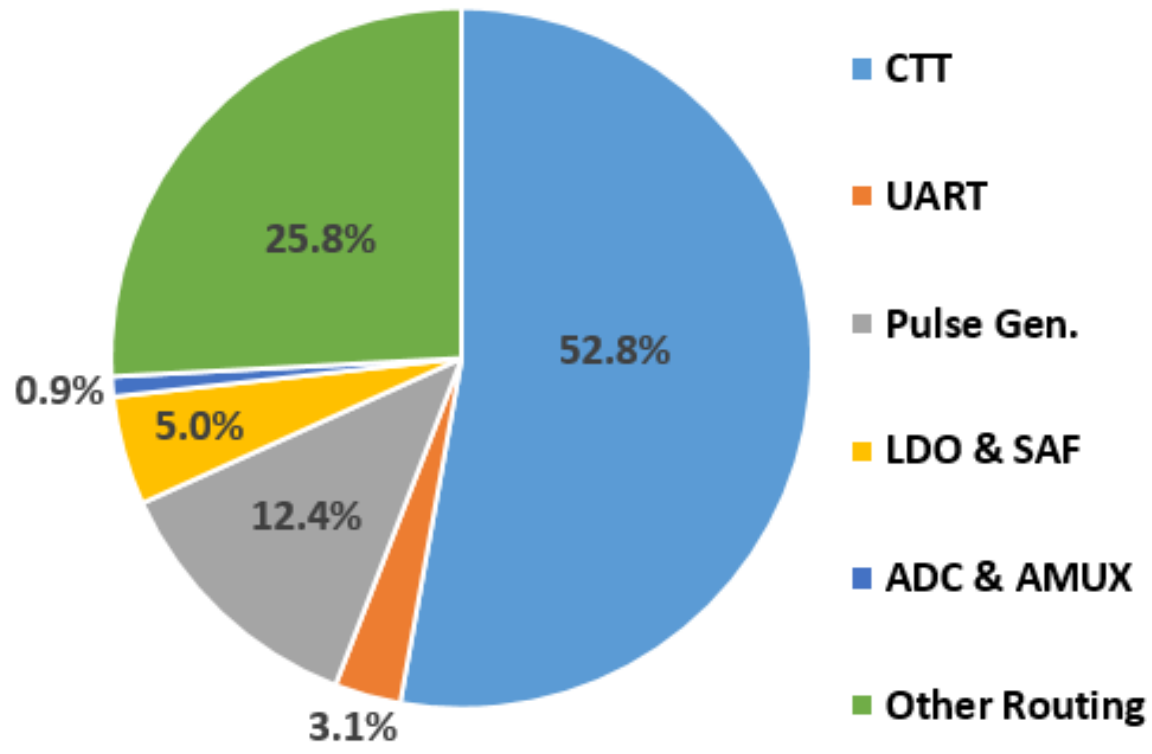

Fig. 11. Area breakdown of the CTT engine

Table I compares the CTT engine with pure digital computing engine in terms of process, area, power, clock speed, peak MAC numbers, etc. CTT-based analog computing engine occupies around 1/10 area to provide more than 500 times computation resources. Compared to the memristor-based design in [33], the CTT engine also consumes about 2x power but it should be noted that the clock speed is running at 10x faster while being more CMOS-compatible. Reference [40] has much lower power consumption because it operates in sub-threshold region and aims for super low-speed applications, but the CTTs' density is much higher than that in [40] even after normalized by the technology nodes.

TABLE I
PERFORMANCE SUMMARY

| Merits | Memristor-Based[33] | Memristor-Based[40] | Digital-Based[30] | This work |
|---|---|---|---|---|
| Process | 90 nm CMOS | 130 nm SONOS | 28 nm CMOS | 28 nm CMOS |
| Core Area ($mm^2$) | 1.86/0.74(3) | 2.41(2) | 5.8 | 0.68(4) |
| Power (mW) | 6.45 | 0.0009(2) | 41 | 14.8(4) |
| Clock Speed(MHz) | 50 | -- | 200-1175 | 500 |
| Stored Weights# | 65.5 K | 14 K | -- | 614K |
| Normalized Synapses Area(1) | 3.5K/1.4K | 10 K | -- | 1.4K |
| MACs # | -- | -- | 0.64 K | 76.8K(5) |
| SRAM Size | 0 | 0 | 128KB | 0 |
| Non-Volatile | Yes | Yes | No | Yes |
| CMOS Compatible | Yes | No | Yes | Yes |

(1) Normalized synapses area is calculated from core area / synaptic weights# / feature size of technology node$^2$
(2) Not include ADC
(3) Estimated for memory crossbar, ADCs and pulse generation circuits
(4) Only include neural network computing engine
(5) Equivalent MACs # is reduce by 8 times due to SAF.

## VII. CONCLUSION

We have demonstrated that the CTT, as a fully-CMOS-compatible non-volatile analog device, enables an analog computing engine to calculate fully connected neural networks. The proposed architecture with the mixed-signal analog-digital interfaces supports single and multi-layer fully connected neural networks' computation, and the inter-layer partial calculation results can be flexibly stored in any type of available memory and processed with any calibration and optimization algorithms to guarantee analog computing accuracy. A 784 x 784 CTT array is used for handwritten digit recognition problem and more than 95% accuracy is achieved with the 8-bit fixed-point analog-digital interface.

Finally, a physical design is provided using standard TSMC HPM 28nm PDK to estimate area and power consumption. Since high-k gate dielectrics are expected to be present in all current and future CMOS technology nodes, the integration of the proposed architecture with other functional components should be seamless. Future work of this paper includes multilayer networks and training algorithms, considering non-idealities such as nonlinearity, process variations, etc. The findings of this paper pave the way for an ultra-large scale, low power, low cost and high-performance fully CMOS-compatible intelligent system.

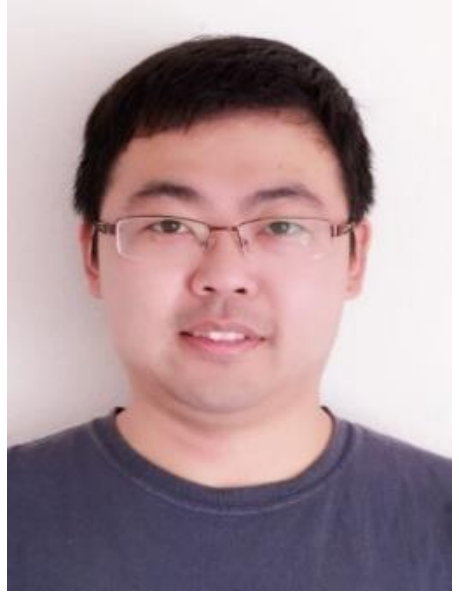

**Yuan Du** (S'14-M'17) received his B.S. degree with honor in electrical engineering from Southeast University (SEU), Nanjing, China, in 2009, his M.S. and his Ph.D. degree both in electrical engineering from University of California, Los Angeles (UCLA), in 2012 and 2016, respectively. His research interests include designs of machine-learning hardware accelerator, high-speed wireline/SerDes, and RFICs. He was the recipient of the Microsoft Research Asia Young Fellowship (2008), Southeast University Chancellor's Award (2009) and Broadcom Fellowship (2015).

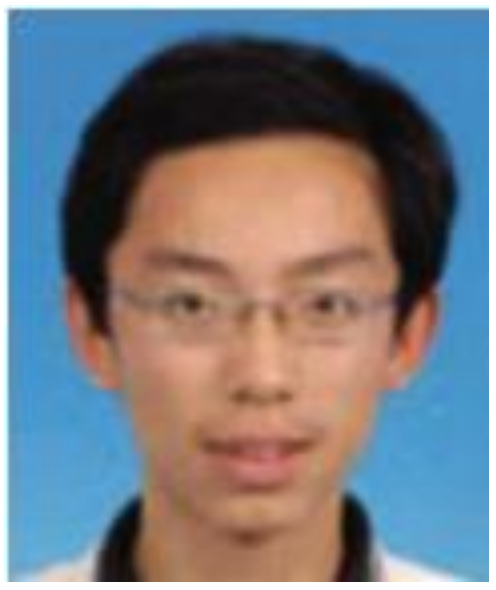

**Li Du** (M'16) received the B.S. in Information Science and Engineering from Southeast University, Nanjing, China, in 2011. He later became an MS student, majoring in Electrical Engineering, at the University of California, Los Angeles (UCLA). At UCLA, he worked in the High-Speed Electronics Lab and was in charge of designing high-performance mixed-signal circuits for communication and touch-screen systems. From June 2012 to October 2012 he worked as an intern in the Broadcom Corporation FM radio team, in charge of designing a second order continuous-time delta-sigma ADC for directly sampling FM radios. From June 2013 to Sept 2016, he works at Qualcomm Inc, designing mixed signal circuits for cellular communications. He received his Ph.D from UCLA in 2016 and is a hardware architect in Kneron Inc. now.

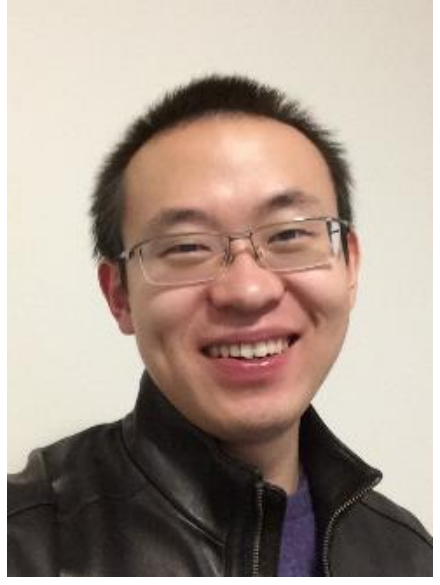

**Xuefeng Gu** received the B.E. in Information Engineering at Southeast University and M.S. in Electrical Engineering at the University of California, Los Angeles, in 2011 and 2013, respectively. He is currently pursuing a Ph.D. in Electrical Engineering at UCLA. His research mainly focuses on the characterization and modeling of charge-trap transistors and their application in neuromorphic computing systems. He was previously with IBM working on Product Technology Interaction in 14-nm FinFET technology.

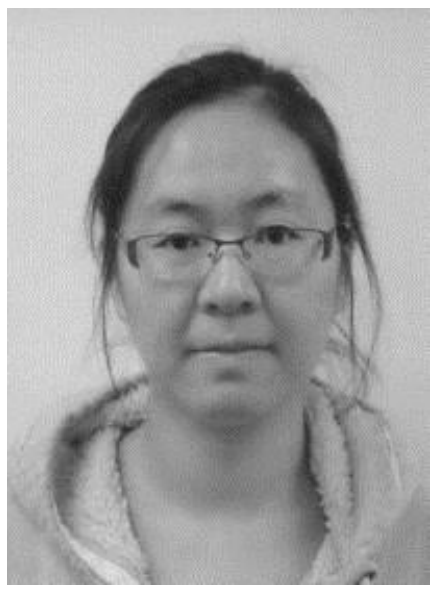

**Jieqiong Du** received the B.S. degree in microelectronics from Shanghai Jiao Tong University, China, in 2012, and the M.S. degree in electrical engineering from University of California, Los Angeles (UCLA), in 2014. She is currently pursuing the Ph.D. degree in electrical engineering from University of California, Los Angeles (UCLA). Her research interests include high-speed mixed signal circuits and I/O links.

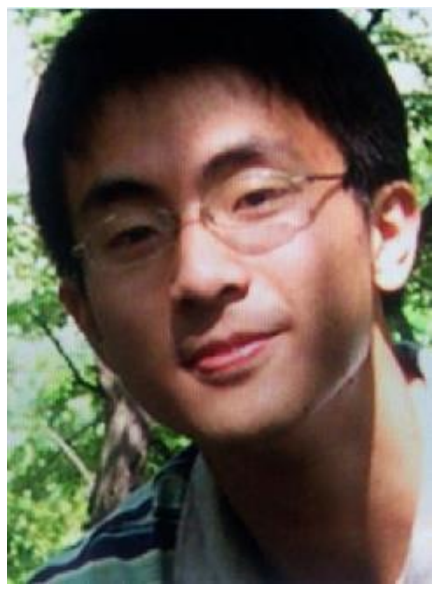

**X. Shawn Wang** received the B.S. degree in electrical engineering at University of Electronic Science and Technology of China (UESTC), Chengdu, China, in 2011, the M.S. degree in electrical and computer engineering at the University of California, Santa Barbara (UCSB), CA, USA, in 2013, and the Ph.D. degree in electrical and computer engineering at the University of California, Los Angeles (UCLA), CA, USA, in 2018. He is a mixed-signal designer at Qualcomm Inc, CA, USA. His research interests are mixed-signal and RFIC designs.

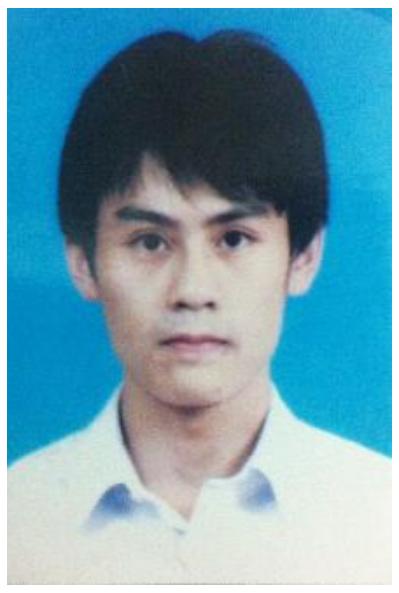

**Boyu Hu** received the B.Sc degree (Hons.) from Chu-Ko-Chen Honors College, Zhejiang University, Hangzhou, China in 2008, the M.S degree from Institute of VLSI Design, Zhejiang University, Hangzhou, China, in 2011, and Ph.D degree from University of California, Los Angeles, in 2016. His research interests include high-speed/high-precision mixed signal integrated circuit and system; hardware mapping of complex signal processing algorithms and its related VLSI architecture and design. He is a recipient of UCLA EE department fellowship and Broadcom fondation fellowship.

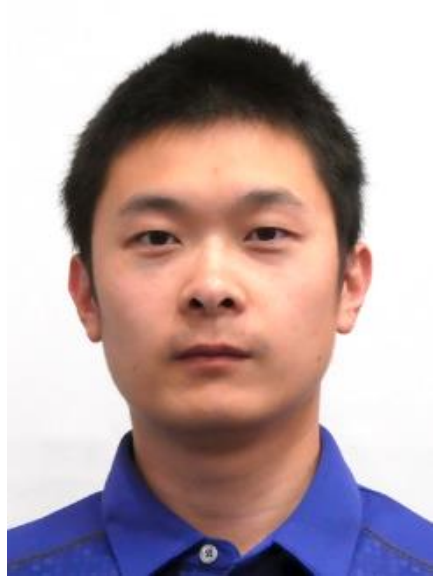

**Mingzhe Jiang** received the B.S. degree from the Department of Computer Science and Engineering at the Shanghai University of Electric Power, Shanghai, China in 2010 and received M.S. degree in electrical engineering from University of California, Los Angeles in 2012. Since 2012, he joined Maxim Integrated Products as digital IC deisgn engineer and contributed to several important power management IC products used in cell phone market. In 2017, he joined Kneron Inc as senior hardware architect for deep learning accelerator. His research interests focused on deep learning accelerator architecture, structure light 3D reconstruction and face recognition using convolution neural network.

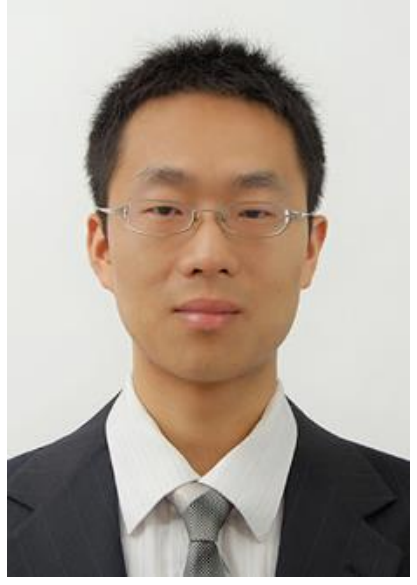

**Xiaoliang Chen** received the B.S. degree in electrical engineering from Tsinghua University, Beijing, China, in 2007 and the M.S. degree in computer science from Peking University, Beijing, China in 2010. He is currently pursuing a Ph.D. degree in computer engineering at University of California, Irvine, USA. His current research interests include approximate computing and machine-learning hardware accelerator design. Since 2013 he has been with Broadcom Inc., Irvine, CA, USA, developing design flow and methodology for analog and mixed-signal circuits.

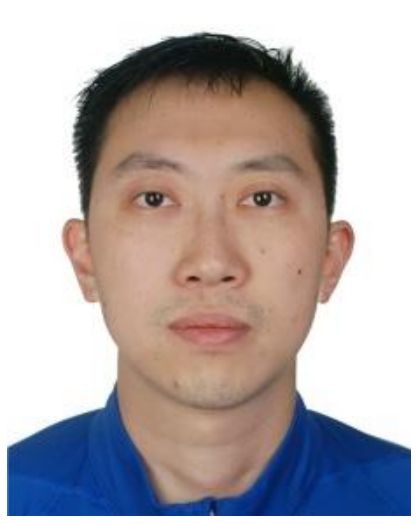

**Junjie Su** received the B.S. in Electrical Engineering and Computer Science from University of California, Berkeley, in 2006. He later received his M.S. in Electrical and Computer Engineering from University of California, San Diego in 2008. He has more than 7 years ASIC design and verification industry experience working in Marvell, Broadcom, and Synopsys. His expertise is in Heterogeneous Computing and Embedded System. He is now a principal A.I. research scientist in Kneron Inc.

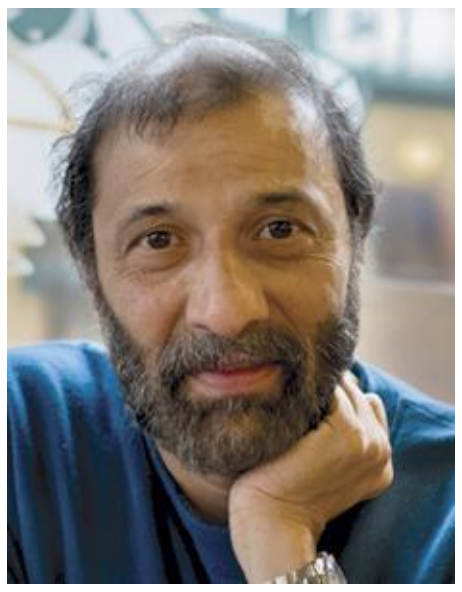

**Subramanian S. Iyer** (F'95) is Distinguished Professor and holds the Charles P. Reames Endowed Chair in the Electrical Engineering Department, University of California, Los Angeles, CA, USA, and the Director of the Center for Heterogeneous Integration and Performance. His key technical contributions have been the development of the world's first SiGe base HBT, salicide, electrical fuses, embedded DRAM, and 45-nm technology node. He also was among the first to commercialize bonded SOI for CMOS applications through a startup called SiBond LLC. He has authored over 300 papers and holds over 70 patents. His current research interests include advanced packaging and 3-D integration for systemlevel scaling and new integration and computing paradigms, as well as the long-term semiconductor and packaging roadmap for logic, memory, and other devices including hardware security and supply chain integrity. Dr. Iyer was a recipient of several Outstanding Technical Achievements and Corporate Awards at IBM. He is an APS Fellow, IBM Fellow, and a Distinguished Lecturer of the IEEE EDS as well as its treasurer and a member of the Board of Governors of the IEEE EPS. He is also a fellow of the National Academy of Inventors. He is a Distinguished Alumnus of IIT Bombay and also a recipient of the IEEE Daniel Noble Medal for emerging technologies in 2012.

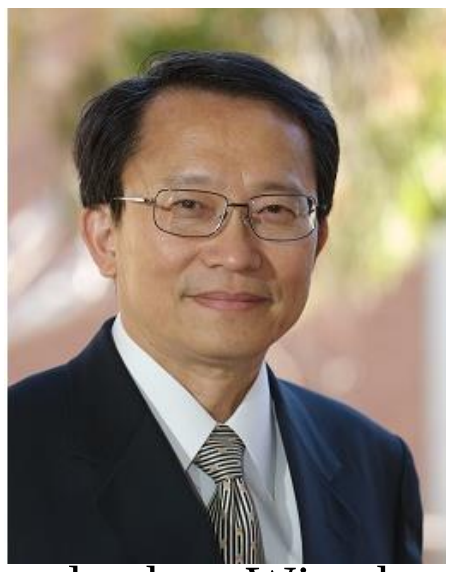

**Mau-Chung Frank Chang** (M'79–SM'94–F'96-LF'17) received the B.S. degree from National Taiwan University (1972), the M.S. degree from National Tsing Hua University (1974), and the Ph.D. from National Chiao Tung University (1979). He is currently the President of National Chiao Tung University (NCTU), Hsinchu, Taiwan and the Wintek Distinguished Professor of Electrical Engineering at the University of California, Los Angeles (UCLA), CA, USA. He also served as the Chair of the EE Department from 2010 to 2015. Before joining UCLA, he was the Assistant Director of the High Speed Electronics Laboratory of Rockwell International Science Center (1983–1997), Thousand Oaks, CA, USA. In this tenure, he developed and transferred the Heterojunction Bipolar Transistor (HBT) integrated circuit technologies from the research laboratory to the production line (later became Skyworks). The HBT productions have grown into multi-billion dollar businesses and dominated the cell phone power amplifier and front-end module markets for the past twenty years (currently exceeding 10 billion-units/year and 50 billion units in the last decade). Throughout his career, Dr. Chang has pursued his research in areas of high-speed electronics, integrated circuit/system designs for radio, radar and imagers, and multiband interconnects for intra- and inter-chip communications. Dr. Chang is recognized by his memberships with the U.S. National Academy of Engineering (2008) and Academia Sinica of Taiwan (2012). He is a Fellow of IEEE (1996) and received David Sarnoff Award in 2006.